\documentclass[prb,aps,twocolumn,groupedaddress,floats,showpacs,final,superscriptaddress]{revtex4-2}
\usepackage[latin3]{inputenc}
\usepackage[makeroom]{cancel}
\usepackage{graphicx}
\usepackage{amsmath}
\usepackage{amsfonts}
\usepackage{amssymb}
\usepackage{chngcntr}
\usepackage{color}
\usepackage[left=2cm,right=2cm,top=2cm,bottom=2cm]{geometry}

\usepackage{graphicx} 
\usepackage{dcolumn}
\usepackage{bm}
\usepackage{simplewick}
\usepackage{array}
\usepackage{appendix}

\RequirePackage[
   hyperindex,colorlinks,bookmarksnumbered,
   plainpages=true,pdfstartview=FitH]{hyperref}
\hypersetup{linkcolor=blue,urlcolor=blue,citecolor=blue}
\usepackage{hyperref}
\usepackage{mathrsfs}
\definecolor{purple}{rgb}{0.5,0,0.6}

\usepackage{ulem}
\renewcommand{\emph}[1]{\textit{#1}}
\definecolor{darkblue}{rgb}{0,0,0.5}
\definecolor{darkgreen}{rgb}{0,0.5,0}
\definecolor{darkred}{rgb}{.7,0,0}
\definecolor{purple}{rgb}{0.5,0,0.6}
\definecolor{orange}{rgb}{1,0.5,0}
\definecolor{grey}{rgb}{.6,.6,.6}
\definecolor{lightpink}{rgb}{1,0.7,0.75}
\definecolor{pink}{rgb}{1,0.4,0.58}
\definecolor{deeppink}{rgb}{1,0.08,0.58}

\usepackage{ulem}

\begin{document}

\date{\today}
\title{Manifestation of Luttinger liquid effects in a hybrid metal-semiconductor double-quantum dot device}

\author{A. V. Parafilo}
\email{aparafil@ibs.re.kr}
\affiliation{Center for Theoretical Physics of Complex Systems, Institute for Basic Science, Expo-ro 55, Yuseong-gu, Daejeon 34126, Republic of Korea}

\date{\today}

\begin{abstract}
We theoretically study the transport properties of a hybrid nanodevice comprised of two large metallic islands incorporated in a two-dimensional electron gas. The high-tunability of the conducting channels electrically connecting two islands to each other and to the leads allows us to treat the setup as a realization of a multi-channel two-site charge Kondo (2SCK) model. It is shown that the leading temperature dependence of the conductance in the 2SCK circuit satisfies the conductance scaling of a single-impurity problem in a Luttinger liquid, whose interaction parameter is fully determined by the number of conducting channels in the device. 
We demonstrate that the finite weak backscattering in all conducting channels features the appearance of the subleading temperature dependencies in linear conductance. At the special critical point, we predict an equivalency between the 2SCK nanodevice and a single-site two-channel charge Kondo problem, where one Kondo channel is implemented by a non-interacting electron gas and the second Kondo channel is attributed to the Luttinger liquid.
\end{abstract}

\maketitle

\section{Introduction}

Recent advances in the fabrication and control of hybrid metal-semiconductor nanodevices have induced enormous interest among the theoretical and experimental community due to their unprecedented access to the multichannel Kondo physics and promising perspective as a platform for studying quantum critical phenomena. 

Although the original Kondo model \cite{kondo} is usually attributed to the interaction between spins of conduction electrons and impurity spins~\cite{hewson}, Kondo physics may arise whenever degenerate quantum states are coupled to one or few bath continua~\cite{legget,lehur2,lehur_review}. A 'charge' implementation of the Kondo model, suggested in seminal papers~\cite{matveev1,flensberg,matveev2,furusakimatveev}, consists of a large quantum dot (QD) in the weak Coulomb blockade regime attached to one or a few leads via high-controllable quantum point contacts (QPCs). Two degenerate macroscopic charge states of the QD tuned by a gate voltage constitute impurity pseudo-spin, and the electron's location (whether it is in the lead or in the QD) plays the role of the conduction electron's pseudo-spin. The corresponding mapping of the single-electron transistor model on the anisotropic Kondo model is justified at $ T\ll E_C$ (where $E_C$ is the QD's charging energy) and if the QD has a continuous density of states. The 'charge' Kondo (CK) model can be easily generalized to the multi-channel Kondo model, where the number of Kondo channels is determined either by the electron's internal degree of freedom (spin), the number of electronic channels or the number of electrodes coupled to the QD.  

The first experimental implementation of the CK device was developed in a hybrid metal-semiconductor single-electron transistor on the base of a Ga(Al)As heterostructure operating in the integer quantum Hall regime~\cite{pierre1, pierre3}. A semiconductor component of the setup ensures the high tunability of electric channels, while a metallic island with negligible level spacing provides the quantization of charge states and avoids any coherent electronic transport from a reservoir on one side of the island to another. As a consequence, the universal conductance scalings for the cases of two- and three-channel CK models were experimentally reported with unprecedented control~\cite{pierre1, pierre3}. It is essential to mention that hybrid CK circuit~\cite{pierre1, pierre3} can also be regarded as a simulator for the metal-to-insulator transition occurred in 1D interacting gas with a single-impurity~\cite{pierre,DCB}. Indeed, a quantum channel with arbitrary transmission (implemented by the QPC) being in series with linear resistance determined by $N$ ballistic channels corresponds to the problem of the Luttinger liquid (LL) of interaction parameter $K=N/(N+1)$ with an isolated impurity~\cite{devoret, safisaleur}.

Nevertheless, the interest generated by the experiments~\cite{pierre1, pierre3} has resulted in a number of exciting experimental~\cite{heatcoulomb, pierre, inter, DCB, DGG, pierre4, pierre5}  and theoretical works~\cite{mitchel,thanh2018,thanh2024,karkinew, VN2, thanhprl,VN, parafilo, parafilo2, parafilo3, parafilo4, idrisov, idrisov2, sim, sim2, sela1, sela2, fritz, karki1, karki2, karki3, kiselev, parafilonew, florian}. Especially, we would like to emphasize Ref.~\cite{thanh2018}, where a so-called two-site CK (2SCK) nanodevice was first proposed to test competing between the Fermi- and non-Fermi liquid phases arising in the setup.
Authors in Ref.~\cite{thanh2018} discussed a double-quantum dot (DQD) device with strong- and weak inter-dot coupling, where each dot is attached to a fixed number of leads via single-mode QPCs.
Therefore, the theoretical proposal in~\cite{thanh2018} established a 'charge' implementation of a quantum two-impurity model, where the interplay between an individual (multi-channel) Kondo screening and Ruderman-Kittel-Kasuya-Yosida-like inter-dot interaction may take place. 

Soon, introduced in~\cite{thanh2018} model was fabricated in a hybrid metal-semiconductor double-island device~\cite{DGG} as a platform for probing frustrated interaction at the exotic quantum critical point with fractional excitation~\cite{karki1, karki2}. In the simplest case of strong inter-dot coupling, when each QD is coupled only to one electrode via nearly ballistic single-mode QPC, the emergence of a $\mathbb{Z}_3$ parafermion featured by the fractional $\log(3)/2$ residual entropy was declared~\cite{DGG, karki2}. 

Further, it was revealed in Ref.~\cite{parafilonew} that the 2SCK nanodevice, where left/right QD connected via $N$($M$) fully ballistic channels to source/drain reservoirs, can be treated as a LL simulator with the Luttinger interaction parameter $K$$=$$NM/(N+M+NM)$. Here, the interaction parameter $K$ may acquire a value in the interval $1/3<K<1$. The study in Ref.~\cite{parafilonew} didn't account for finite backscattering in the QPCs connecting QDs with source and drain leads (except for a special case $N$$=$$M$$=$$1$). Thus, the possible interplay between interaction effects and quantum CK criticality was neglected therein. Meanwhile, such interplay may produce exciting phenomena involving various quantum phase transitions and crossovers~\cite{baranger1,baranger2,baranger3}; see also a series of works~\cite{VN, parafilo, parafilo3,parafilo2,parafilo4}, where the influence of interactions on the transport properties in a single-site CK circuit was investigated.

This paper partially corrects this omission by considering the 2SCK device with strong inter-dot coupling and includes the finite backscattering in all QPCs.

\begin{figure}
\centering 
\includegraphics[width=1\columnwidth]{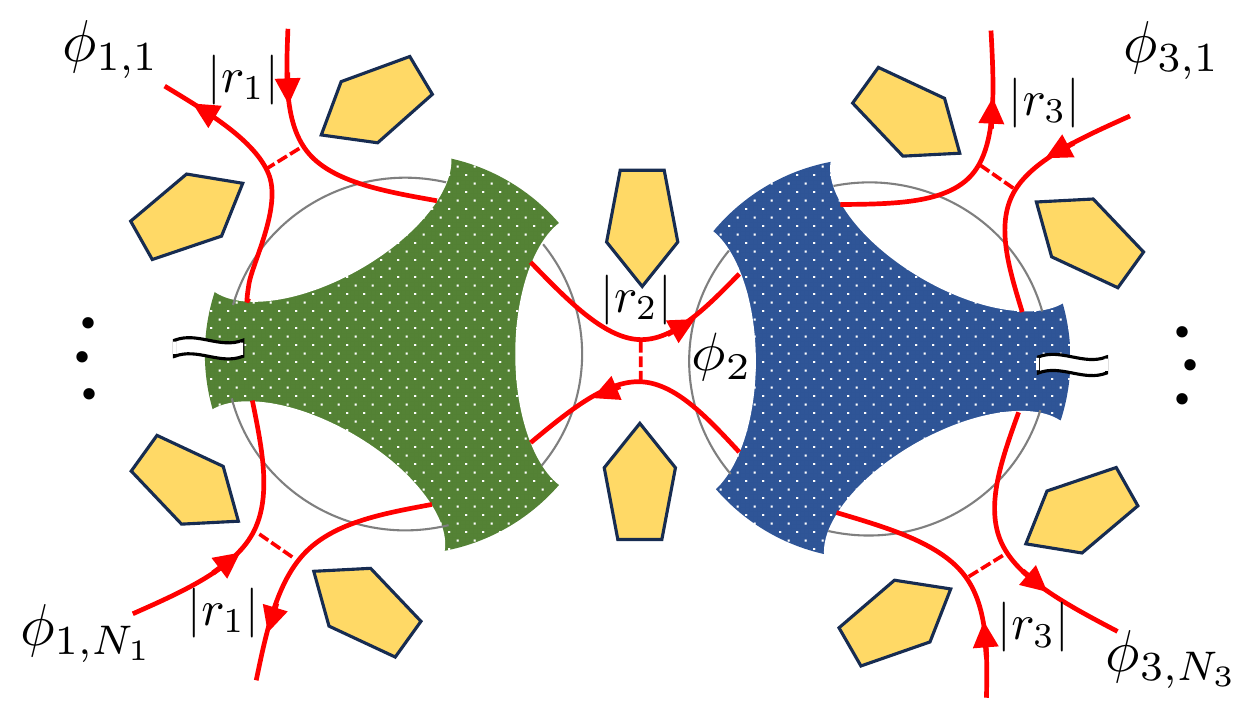} \caption {Schematic representation of the two-site charge Kondo (2SCK) circuit: a hybrid metal-semiconductor device 
formed in Ga(Al)As heterostructure
consists of two metallic islands or QDs with the continuous density of states (marked by green and blue colors) strongly coupled via a single-mode quantum point contact (QPC) to each other. Besides, left (right) QD is electrically connected to the 2DEG via $N_1$~($N_3$) single-mode QPCs.  
The system is exposed to a strong magnetic field, which drives it in the integer quantum Hall regime with filling factor $\nu=1$. Thus, electric current propagates along the spin-polarized edge channels (marked by red solid lines). The single-mode QPCs are assumed to be nearly transparent (featured by small reflection amplitude $|r_{\alpha}|\ll1$ with $\alpha=1,2,3$), resulting in weak backscattering between left- and right-moving electrons. Charge degree of freedom is characterized via bosonic fields $\phi_{\alpha,i}(x)$ (here, $i=1,...,N_{\alpha}$), which is the 'difference' between chiral left- and right-moving bosonic fields, $\phi_{\alpha,i}=(\varphi^L_{\alpha,i}-\varphi^R_{\alpha,i})/2$~\cite{giamarchi}.}
\label{Fig1} 
\end{figure}

\section{ Model and Effective Hamiltonian} 
We study a multichannel 2SCK circuit -- a hybrid metal-semiconductor device consisting of two large metallic islands (QDs) with a continuous density of states incorporated into a high-mobility Ga(Al)As two-dimensional electron gas (2DEG), see Fig.~\ref{Fig1}. 
The system is in a strong magnetic field, thus realizing the regime of integer quantum Hall effect with a filling factor $\nu=1$. Two QDs are strongly coupled to each other via single-mode QPC, while the left (right) QD is strongly connected to source (drain) electrodes via $N_1$($N_3$) single-mode QPCs. The left- and right-moving (chiral) electrons in the 1D quantum Hall edge channels undergo weak backscattering in each QPCs. To describe the transport properties of the 1D channels, we use bosonic operators $\phi_{\alpha,i}(x)$ and $\theta_{\alpha,i}(x)$, which satisfy the following commutation relation $[\phi_{\alpha,i}(x),\partial_x \theta_{\alpha',j}(x')]=i\pi\delta(x-x')\delta_{\alpha\alpha'}\delta_{ij}$. 
Here, field $\phi_{2,i}(x)$ is responsible for inter-dot connection, while fields $\phi_{1(3),i}$ characterize other $N_1$($N_3$) conducting channels, see Fig.~\ref{Fig1}.
In this representation, the Hamiltonian of edge states is simply the Hamiltonian of the spinless Luttinger liquid with interaction parameter $K=1$~\cite{giamarchi,gogolin}.

The whole system presented in Fig.~\ref{Fig1} can be described by the total Hamiltonian $H=H_0+H_C+H_{BS}$, where (in $\hbar=k_B=1$ units)
\begin{eqnarray}\label{ham0}
&&H_{0}=\frac{v_F}{2\pi}\sum_{\alpha=1}^3\sum_{i=1}^{N_{\alpha}}\int dx \left\{[\pi \Pi_{\alpha,i}(x)]^2+\left[\partial_x\phi_{\alpha,i}(x)\right]^2\right\}\nonumber\\
\end{eqnarray}
is the kinetic energy of conducting channels, $v_F$ is the Fermi velocity and $\Pi_{\alpha,i}(x)$ is the canonically conjugated to the field $\phi_{\alpha,i}(x)$ momentum, $[\phi_{\alpha,i}(x),\Pi_{\alpha',i'}(x')]=i\delta(x-x')\delta_{\alpha\alpha'}\delta_{ii'}$. Since for the considered model $N_2=1$, we use the notation $\phi_{2,1}(x)\equiv \phi_2(x)$ in what follows.

Two metallic islands in the weak Coulomb blockade regime can be described via the following Hamiltonian:
\begin{eqnarray}
&& H_C=\frac{E_C}{\pi^2}\left(\sum_{i=1}^{N_1}\phi_{1,i}(0)-\phi_2(0)+\pi \mathcal{N}_{g}\right)^2\nonumber\\
&&~~~~~~~+\frac{E_C}{\pi^2}\left(\phi_2(0)-\sum_{i=1}^{N_3}\phi_{3,i}(0)+\pi \mathcal{N}_{g}\right)^2,
\end{eqnarray}
where $E_C=e^2/2C$ is the charging energy of islands, $C$ is islands' capacitance; $\mathcal{N}_{g}$ is a gate voltage (assumed being equal for both dots). 

A weak scattering between right- and left-moving particles in QPCs is characterized by
\begin{eqnarray}
&& H_{BS}= \sum_{\alpha=1}^3\sum_{i=1}^{N_{\alpha}}\frac{D |r_{\alpha,i}|}{\pi}\cos\left[2\phi_{\alpha,i}(0)\right].
\end{eqnarray}
Here, $D$ is an ultra-violet energy cut-off associated with the existence of bandwidth, and $|r_{\alpha,i}|\ll 1$ are the backscattering matrix elements. We consider the case when all backscattering amplitudes in QPCs with the same index $\alpha$ are equal to each other, $|r_{\alpha,i}|\equiv |r_{\alpha}|$.

Next, we perform few approximations to simplify introduced model. First, we present $\phi_{1,i}$ ($\phi_{3,i}$) via new fields $\phi_{1,c}, \phi_{1,s}, ... , \phi_{1,f}$ ($\phi_{3,c}, \phi_{3,s}, ... , \phi_{3,f}$), which characterize the charge, spin and other flavor modes. With this representation only three fields, namely, two charge modes $\phi_{\alpha,c}=\sum_{i=1}^{N_{\alpha}}\phi_{\alpha,i}/\sqrt{N_{\alpha}}$ ($\alpha=1,3$) and $\phi_{2}$, appear in the Coulomb blockade Hamiltonian. Second, we perform a unitary transformation to diagonalize $H_0+H_C$ part of the total Hamiltonian $H$, which reduces the fields $(\phi_{1,c},\phi_{2},\phi_{3,c})$ to $(\phi_{A},\phi_{B},\phi_{C})$ as follows:  
\begin{eqnarray}\label{vector}
\left(\begin{array}{c}
\phi_{1,c} \\ \phi_2 \\ \phi_{3,c}
\end{array}\right)=\hat P\left(\begin{array}{c}
\phi_A \\ \phi_B \\ \phi_C
\end{array}\right) +\pi\left(\begin{array}{c}
-\frac{\mathcal{N}_{g}}{\sqrt{N_1}} \\ 0 \\ \frac{\mathcal{N}_{g}}{\sqrt{N_3}}
\end{array}\right).
\end{eqnarray}
An exact expression for the matrix $\hat P$ can be found in Appendix~\ref{appA}. 
After diagonalization, the kinetic energy Hamiltonian $H'_0=\hat P^{-1}H_0\hat P$ has the same quadratic form as in Eq.~(\ref{ham0}) (just written in terms of new variables), while the Coulomb blockade Hamiltonian acquires a more simple form 
\begin{eqnarray}
&& H'_{C} = \frac{ E_{C}}{\pi^2}\left\{\mathcal{M}_B\,\phi_{B}^{2}(0)+\mathcal{M}_C\,\phi_{C}^{2}(0)\right\}.
\end{eqnarray}
Here, $\mathcal{M}_B$, $\mathcal{M}_C$ are effective mass parameters of fields $\phi_B$, $\phi_C$, whose values are determined by the number of total conducting channels in the system as 
\begin{eqnarray}
\mathcal{M}_{B(C)}=1+\frac{(N_1+N_3)}{2}\mp\sqrt{1+\frac{(N_1-N_3)^2}{4}}.
\end{eqnarray}
Thus, one obtains an effective Hamiltonian for $N_1+N_3+1$ modes, two of which $(\phi_B,\phi_C)$ are gapped due to the Coulomb blockade and $N_1+N_3-1$ are massless.  

In this paper, we calculate the linear conductance through the device in zero-frequency limit using the Kubo formula and current operator $\hat I=-(e/\pi)\sqrt{N_1}\partial_t \phi_{1,c}(0,t)$, see, e.g., \cite{furusakimatveev,DCB}. In terms of new variables $\phi_{\alpha}$ ($\alpha=A,B,C$), it reads
\begin{eqnarray}\label{kubo}
G=K G_{0}\frac{2T}{\pi i}\lim_{\omega\rightarrow0}\omega \lim_{i\omega_n\rightarrow \omega +i0^{+}}\langle \phi_A(i\omega_n)\phi_A(-i\omega_n)\rangle,
\end{eqnarray}
where $G_0=e^2/2\pi$ is a unitary conductance, $ K = N_1N_3/(N_1+N_3+N_1N_3)$ is an effective LL interaction parameter (see discussion below), and $\phi_{A}(i\omega_n)=\int_0^{T^{-1}}d\tau \phi_{A}(\tau)\exp(i\omega_n\tau)$, while $\omega_n=2\pi T n$ is the Matsubara frequency and $\tau$ is the imaginary time. The thermal average $\langle \phi_A(i\omega_n)\phi_A(-i\omega_n)\rangle$ can be calculated by utilizing the functional integration technique and second-order perturbation theory over $|r_{\alpha}|\ll1$. It is important that only one gapless mode $\phi_A(x)$ out of $N_1+N_3-1$ determines the charge transport through the 2SCK setup.

Further simplification of the model is possible since the charging energy suppresses the fluctuations of $\phi_B$, $\phi_C$ at low temperatures $T\ll E_C$. Thus, one can integrate out these modes by replacing the backscattering Hamiltonian $H'_{BS}$ with its value averaged over the fluctuations of $\phi_B$, $\phi_C$. As a result, one obtains effective Hamiltonian, which, since we are interested in $\phi_A(x)$ mode, can be written as 
\begin{eqnarray}\label{effective}
H_{\rm eff}= H_{\rm SG}[\phi_A]+\sum_{j=1}^{N_1+N_3-2}H_{\rm 1d}[\phi_{j}]+H_{\rm int}[\phi_A,\phi_j].\nonumber\\
\end{eqnarray}
Here,
\begin{eqnarray}\label{SG}
H_{\rm SG}[\phi_A]=\frac{v_F}{2\pi}\int dx \left\{[\pi \Pi_A(x)]^2+[\partial_x\phi_A(x)]^2\right\}\nonumber\\+\beta \cos\left[2\sqrt{K}\phi_A(0)\right]
\end{eqnarray}
is a, so-called, boundary sine-Gordon Hamiltonian, which describes many different physical systems including a single-impurity scattering problem in the LL with interaction $K$ \cite{kanefisher,giamarchi,gogolin}, conducting channel coupled to an Ohmic environment \cite{safisaleur,weiss}, and other~\cite{chakra,shmidt,QBM,grabert1}. Coefficient $\beta$ is a function of $N_1$, $N_3$ as it is shown in Ref.~\cite{parafilonew}, see Eq.~(23) in Supplemental Material therein. 

Effective model (\ref{effective}) becomes non-trivial if $N_1+N_3>2$. In this case, field $\phi_A$ interacts with other massless fields, namely, with $N_1-1$ flavor modes in the left part of the device and with $N_3-1$ flavor modes in the right part of the device. The corresponding Hamiltonian of $j$th copy of 1D massless quasiparticles and interaction term can be written as:
\begin{eqnarray}
H_{\rm 1d}[\phi_j]=\frac{v_F}{2\pi}\int dx \left\{[\pi \Pi_j(x)]^2+[\partial_x\phi_j(x)]^2\right\},\\
H_{\rm int}=\left\{\sum_{i=1}^{N_1-1}\lambda_i\cos [a_i\phi_A(0)]\prod_{j=1}^i\cos[a_{ij}\phi_j(0)]\right.\nonumber\\\left.+
\sum_{i=1}^{N_3-1}\lambda'_i\cos [a'_i\phi_A(0)]\prod_{j=1}^i\cos[a'_{ij}\phi_j(0)]\right\},\label{interaction}
\end{eqnarray} 
where constants $\lambda_i$, $a_i$, $a_{ij}$ and $\lambda'_i$, $a'_i$, $a'_{ij}$ depend on the number of conducting channels $N_1$, $N_3$. In addition, $\lambda_i$, $\lambda'_i$ are determined by the reflection coefficients in QPCs. In general form, Eq.~(\ref{effective}) is too cumbersome. Thus, we concentrate our study on particular cases when $N_1, N_3= \{1, 2, 3\}$. The corresponding effective Hamiltonians Eq.~(\ref{effective}) for these cases in explicit form are shown in Appendix~\ref{appB}.

\section{Results}

\begin{center}
\begin{table}\label{table1}
\begin{tabular}{|| c | c | c | c | c | c | c | c ||}
 \hline
 $N_1$ & $N_3$ & $K$ & $\mathcal{I}_K$ & $g$ & $\mathcal{I}_{g}$ & $g'$ & $\mathcal{I}_{g'}$ \\ [1ex]
 \hline\hline 
 1 & 1 & 1/3 & $T^{-4/3}$ & - & - & - & -
 \\ [1ex]
 \hline
 1 & 2 & 2/5 & $T^{-6/5}$ & 3/5 & $T^{-4/5}$ & - & - \\ [1ex]
 \hline
 1 & 3 & 3/7 & $T^{-8/7}$ & 5/7 & $T^{-4/7}$ & - & - \\ [1ex]
 \hline
 2 & 2 & 1/2 & $T^{-1}$ & 5/8 & $T^{-3/4}$ & - & - \\ [1ex]
 \hline
3 & 3 & 3/5 & $T^{-4/5}$ & 11/15 & $T^{-8/15}$ & - & - \\ [1ex] 
 \hline
 2 & 3 & 6/11 & $T^{-10/11}$ & 7/11 & $T^{-8/11}$ & 8/11 & $T^{-6/11}$ \\ [1ex] 
 \hline
\end{tabular}
\caption{ Effective Luttinger liquid interaction parameters $K$, $g$, $g'$, and corresponding leading, sub- and sub-sub-leading temperature dependencies of the conductance Eq.~(\ref{result}) for different number of conducting channels $N_1$,$N_3=\{1, 2, 3\}$, which couple DQD with source and drain electrodes. For considered in this paper geometry of the system, the Luttinger liquid interaction parameter $K$ is determined as $K=N_1N_3/(N_1+N_3+N_1N_3)$, while $g$($g'$) can be found from Eq.~(\ref{constants}).}
\end{table}
\end{center}

First, let's shortly revise the simplest while the very non-trivial case of $N_1=N_3=1$, which has been already studied in great detail theoretically \cite{karki1,karki2} and experimentally \cite{DGG, karki2}. Since there is only one gapless field $\phi_A$, the 2SCK model with $N_1=N_3=1$ can be described by Eq.~(\ref{SG}) with $K=1/3$ and $\beta$ being the function of gate voltages on both QDs and the backscattering coefficients. The system 'simulates' properties of the LL with $K=1/3$ with a single weak potential barrier. As it is known, the charge transport through such a device is suppressed at low temperatures since $K<1$~\cite{kanefisher,giamarchi,gogolin}. However, cos-term from Eq.~(\ref{SG}) could be nullified by tuning constant $\beta$ with $\mathcal{N}_g$ and $r_{\alpha}$ at, so-called, \textit{triple point}~\cite{karki1,karki2}. As it was shown~\cite{karki1,karki2}, this quantum critical point corresponds to the existence of the fractionalized $\mathbb{Z}_3$ parafermion, which was also identified as a source of $\log(3)/2$ residual entropy. The problem in~\cite{karki2} was solved by utilizing the Emery-Kivelson mapping~\cite{emerykivelson} in the Toulouse limit in full analogy with the two- and four-channel Kondo problems~\cite{gogolin}.  

The same configuration of the 2SCK setup ($N_1=N_3=1$) in the case of the Fractional Quantum Hall regime, when the edge states conduct fractional charge $e^{\ast}=\nu e$ with $\nu=1/m$ ($m$ is an odd integer) being a fractional filling factor, has been investigated in Ref.~\cite{parafilonew}.

Next, we examine the original results. For other than mentioned above particular case $N_1$$=$$N_3$$=$$1$, the linear conductance Eq.~(\ref{kubo}) is obtained in perturbation theory with respect to backscattering coefficients $|r_\alpha|$. We omit here the details of calculations, referring to Refs.~\cite{furusakimatveev,parafilo4,parafilonew} and to Appendix~\ref{appB} for the explicit expressions of used effective Hamiltonians. In general form, the linear conductance through the device reads
\begin{eqnarray}\label{result}
G(T)=KG_0\left[1- \mathcal{C}_l\,\mathcal{I}_{K}(T)-\mathcal{C}_{sl}\,\mathcal{I}_{g}(T)-\mathcal{C}_{ssl}\,\mathcal{I}_{g'}(T)\right],\nonumber\\
\end{eqnarray}
where $K=(N_1^{-1}+1+N_3^{-1})^{-1}$,
\begin{eqnarray}
&&\mathcal{I}_{g}(T)=K\left(\frac{\pi T}{E_C}\right)^{2g-2}\left(\frac{\gamma}{\pi}\right)^{2-2g}\frac{\sqrt{\pi}\Gamma(g)}{2\Gamma\left(g+\frac{1}{2}\right)},
\end{eqnarray}
and $\Gamma(x)$ is the gamma function, $\gamma=\exp(0.577)$ is the Euler constant. 
Explicit values of the interactions constants $K,g,g'$ from Eq.~(\ref{result}) for particular values $N_1$,$N_3$ are shown in the Table~I, while dependent on the reflection coefficients non-universal constants $\mathcal{C}_{l}$, $\mathcal{C}_{sl}$, $\mathcal{C}_{ssl}$ are presented in the Table~II. Due to the usage of perturbation theory, the result Eq.~(\ref{result}) is justified only down to 'crossover' temperatures $T_{cr}^{(\kappa)}\sim E_C \mathcal{C}_{\kappa}^{1/(2-2\kappa)}$ with  $\mathcal{C}_{\kappa}=\{\mathcal{C}_s,\mathcal{C}_{sl},\mathcal{C}_{ssl}\}$ for $\kappa=\{K, g, g'\}$. For even lower temperatures $T\ll T_{cr}^{(\kappa)}$, a consideration in the regime of strong backscattering (tunnel barriers) is required. An exception is the case of $N_1=N_3=2$ ($K=1/2$) when an exact solution is possible due to refermionization of the problem and its equivalency to the resonant model, see~\cite{footnote}.
\begin{center}
\begin{table}\label{table1}
\begin{tabular}{| c | c | c  |}
 \hline
 $N_1$ & $N_3$ & $\mathcal{C}_l$, $\mathcal{C}_{sl}$, $\mathcal{C}_{ssl}$  \\ [1ex]
 \hline\hline 
 1 & 2 & $\mathcal{C}_l=|r_1|^21.11^2+|r_2|^22^2+4.44|r_1||r_2|\cos(2\pi\mathcal{N}_g)$  \\ [1ex]
  & & $\mathcal{C}_{sl}=1.38^2|r_3|^2/2$
\\ [1ex] \hline
 1 & 3 & $\mathcal{C}_l=|r_1|^21.34^2+|r_2|^21.93^2+5.17|r_1||r_2|\cos(2\pi\mathcal{N}_g)$   \\ [1ex]
& & $\mathcal{C}_{sl}=1.39^2|r_3|^2/3$ \\ [1ex]
 \hline
 2 & 2 & $\mathcal{C}_{l}=4 |r_2|^2$  
 \\ [1ex]
 & & $\mathcal{C}_{sl}=|r_1|^2+|r_3|^2$
 \\ [1ex]
 \hline
3 & 3 &  $\mathcal{C}_{l}=1.9^2|r_2|^2$  \\ [1ex] 
& & $\mathcal{C}_{sl}=1.41^2(|r_1|^2+|r_3|^2)/3$ \\ [1ex] 
 \hline
 2 & 3 & $\mathcal{C}_{l}=1.94^2|r_2|^2$ \\ [1ex]  &  & $\mathcal{C}_{sl}=1.4^2|r_1|^2/2$; $\mathcal{C}_{ssl}=1.43^2|r_3|^2/3$  \\ [1ex]
 \hline
\end{tabular}
\caption{ Explicit dependencies of non-universal constants $\mathcal{C}_l$, $\mathcal{C}_{sl}$, $\mathcal{C}_{ssl}$  from Eq.~(\ref{result}) on reflection coefficients at different QPCs and on different numbers of conductance channels $N_1$,$N_3=\{1, 2, 3\}$ in 2SCK device. The origin of numerical constants can be tracked from the particular effective models presented in Appendix~\ref{appB}.}
\end{table}
\end{center}

Equation (\ref{result}) is the first main result of this paper. The first two terms in Eq.~(\ref{result}) give the leading temperature dependence via the 2SCK device, which is the same as for the conductance scaling of the LL with a single weak potential barrier, whose interaction parameter is $K$~\cite{kanefisher,giamarchi,gogolin}. The following correspondence in the multi-channel 2SCK device was predicted in \cite{parafilonew} for the case $|r_1|$$=$$|r_3|$$=$$0$ and could be attributed to a well-known result by Safi and Saleur~\cite{safisaleur}. Indeed, a one-channel coherent conductor coupled in series with the resistance $R$ is equivalent to the impurity problem in the LL with interaction parameter $K=1/(1+r)$, where $r=R/R_0$ is a dimensionless environmental resistance ($R_0=h/e^2$). For the 2SCK device shown in Fig.~\ref{Fig1}, $\phi_2$ field 'scattered' in $\alpha=2$ QPC plays the role of one-channel coherent conductor, while $N_1$($N_3$) channels connected to the first (second) QD can be treated as $N_1$($N_3$) resistors coupled in parallel, $R_{\alpha}=R_0/N_{\alpha}$. 
As a result, all the channels coupled to DQD implement the total dimensionless resistance $r=(R_1+R_3)/R_0$, and thus, $K=1/[1+ N_1^{-1}+N_3^{-1}]$~\cite{parafilonew}. As it is discussed in Ref.~\cite{furusakimatveev}, an alternative explanation of the power-law temperature behavior of the conductance in CK circuits is related to the Anderson orthogonality catastrophe~\cite{anderson}.

Nevertheless, a finite backscattering in $\alpha$th QPCs ($\alpha=1,3$) results in the appearance of the subleading 
temperature dependencies, see third and fourth terms in Eq.~(\ref{result}). From Table~I, we conclude that the temperature scaling of the sub-leading term is associated with the effective interaction constant $g(g')=[N_1N_3+N_{1}(N_3)-1]/[N_1N_3+N_1+N_3]$. If the numbers of channels in $\alpha=1$ and $\alpha=3$ QPCs are different and not equaled to one, $N_1\neq N_3\neq 1$, then the linear conductance has two sub-leading dependencies, see, e.g., the case of $N_1=2$, $N_3=3$ in the Table~I.

\begin{figure}
\centering 
\includegraphics[width=1\columnwidth]{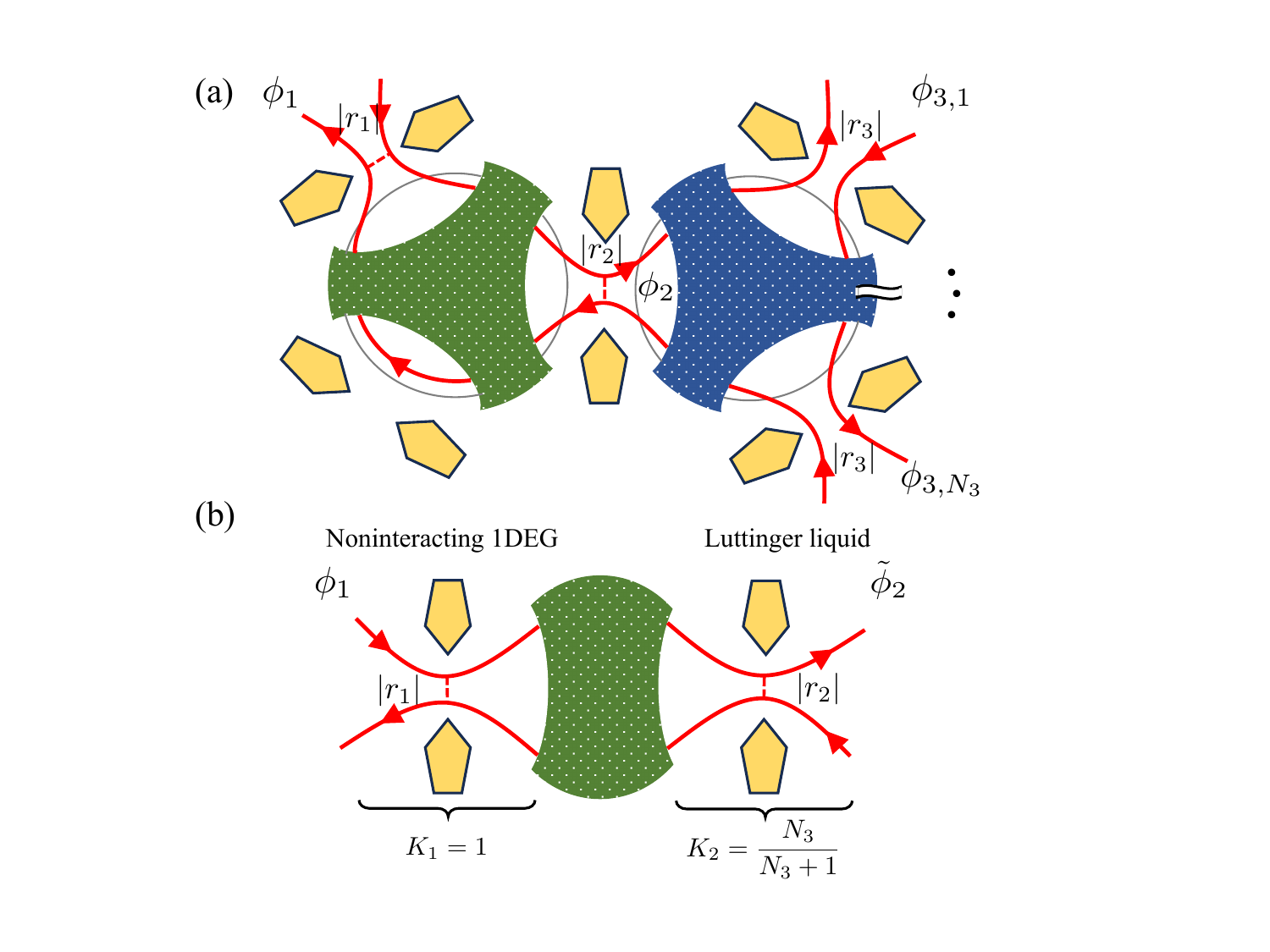} \caption { (a) Two-site charge Kondo (2SCK) nanodevice from Fig.~\ref{Fig1} in a particular case when $N_1$$=$$1$ and $|r_3|=0$. (b)~An equivalent scheme of the above device: a single-site two-channel charge Kondo (2CCK) model, where one channel is attributed to the non-interacting electrons ($K_1=1$) characterized by the bosonic field $\phi_{1,1}(x)\equiv \phi_1(x)$, and the second channel is implemented by the Luttinger liquid (LL) with interaction parameter $K_2=N_3/(N_3+1)$. The following mapping is possible since a conducting channel $\phi_2$ being in a series with linear resistance determined by $N_3$ ballistic channels, $R=R_0/N_3$, is equivalent to the LL with a single impurity. Effective bosonic field $\tilde{\phi}_2(x)$ of the second channel appears after integrating out all fields $\phi_{3,i}(x)$ with $i=1,...,N_3$. A universal Kondo scaling in the vicinity of the 2CCK fixed point reads $G(T)/G_0=(K_{red}/2)[1-(T/T_K)^{K_{red}}]$, where $K_{red}=2K_1K_2/(K_1+K_2)$ is the reduced LL interaction parameter.}
\label{Fig2} 
\end{figure}

Thus, if we ignore for a while the 'interference terms' ($\propto |r_\alpha||r_{\alpha'}|$) and the possible gate-voltage dependence of these terms, one can re-write Eq.~(\ref{result}) in more general form as
\begin{eqnarray}
G(T)=KG_0\left[1-\sum_{\alpha=1}^3 c_\alpha |r_\alpha|^2  \mathcal{I}_{g_{\alpha}}(T)\right],
\end{eqnarray}
where $g_2=K$,
\begin{eqnarray}\label{constants}
&&g_{1(3)}=\frac{N_1N_3+N_{1(3)}-1}{N_1+N_3+N_1N_3},
\end{eqnarray}
and $c_{\alpha}$ are non-universal constants of the order of one. The weak backscattering in $\alpha$th QPC ($\alpha=1,2,3$) leads to its own power-law temperature correction to the maximal conductance.  It's also clear that the subleading term does not appear in case $N_1=N_3=1$ since $g_1=g_2=g_3$. In the same way, the sub-subleading term does not arise for $N_1=1$, $N_3=\{2,3\}$, because $g_1=g_2\neq g_3$.
Nevertheless, tuning the corresponding reflection coefficient can adjust a required temperature dependence in the device. For instance, in the symmetric case $ N_1=N_3\neq1$, the subleading terms become leading ones if the inter-dot QPC is reflectionless ($|r_2|=0$).

However, as one can see from Table II, the 'interference terms' and their gate voltage dependence play an important role. The reason is that tuning the backscattering coefficients $|r_{\alpha}|$ together with $\mathcal{N}_g$ may result in the vanishing of the leading temperature correction. It inspires us to consider a special case when $N_1=1$ (but any $N_3$) and $|r_3|=0$, which allows us to study a quantum critical point associated with the least irrelevant perturbation of the 2SCK model.

\section{Quantum critical point in the case $N_1=1$ and $r_3=0$}

Let's consider the 2SCK circuit when there is only one channel that connects the left QD with 2DEG, $N_1=1$, while all $N_3$ single-mode QPCs that couple the right QD with 2DEG are fully open, $|r_3|=0$, see Fig.~\ref{Fig2}(a). The linear conductance is obtained in a standard way~\cite{furusakimatveev}, within the second-order perturbation theory over the weak backscattering amplitudes $|r_1|$,$|r_2|$ and also accounting for next-to-leading order correction in $T/E_C$: 
\begin{eqnarray}\label{result2}
G(T)=KG_0\left[1-\mathcal{C}_+\left(\frac{\pi T}{E_C}\right)^{2K-2}-\mathcal{C}_-\left(\frac{\pi T}{E_C}\right)^{2K}\right].
\end{eqnarray}
Here, constants $\mathcal{C}_{\pm}$ read as
\begin{eqnarray}
&&\mathcal{C}_+=K\frac{\sqrt{\pi}\Gamma(K)}{2\Gamma(K+1/2)}\left(\frac{\gamma}{\pi}\right)^{2K\frac{1+N_3}{N_3}}r_+^2,\\
&&\mathcal{C}_-=2K\pi^2\frac{\sqrt{\pi}\Gamma(K+1)}{2\Gamma(K+3/2)}\left(\frac{\gamma}{\pi}\right)^{2K\frac{1+N_3}{N_3}}r_-^2,
\end{eqnarray}
while the effective reflection coefficients are
\begin{widetext}

\begin{eqnarray}
&&r_+^2=|r_1|^2\mathcal{M}_B^{2p_{12}^2}\mathcal{M}_C^{2p_{13}^2}+|r_2|^2\mathcal{M}_B^{2p_{22}^2}\mathcal{M}_C^{2p_{23}^2}+2|r_1||r_2|\mathcal{M}_B^{p_{12}^2+p_{22}^2}\mathcal{M}_C^{p_{13}^2+p_{23}^2}\cos(2\pi \mathcal{N}_g),\\
&&r_-^2=|r_1|^2\mathcal{M}_B^{2p_{12}^2}\mathcal{M}_C^{2p_{13}^2}\left[\left(\frac{p_{12}}{\mathcal{M}_B}\right)^2+\left(\frac{p_{13}}{\mathcal{M}_C}\right)^2\right]+|r_2|^2\mathcal{M}_B^{2p_{22}^2}\mathcal{M}_C^{2p_{23}^2}\left[\left(\frac{p_{22}}{\mathcal{M}_B}\right)^2+\left(\frac{p_{23}}{\mathcal{M}_C}\right)^2\right]\nonumber\\&&~~~~~~~~~~~~~+2|r_1||r_2|\mathcal{M}_B^{p_{12}^2+p_{22}^2}\mathcal{M}_C^{p_{13}^2+p_{23}^2}\left[\frac{p_{12}p_{22}}{\mathcal{M}_B^2}+\frac{p_{13}p_{23}}{\mathcal{M}^2_C}\right]\cos(2\pi \mathcal{N}_g).
\end{eqnarray}
\end{widetext}
Explicit expressions for the elements $p_{kl}$ of matrix $\hat P$ are shown in Appendix~\ref{appA}.

The Eq.~(\ref{result2}) is the second main result of this paper. Compared to Eq.~(\ref{result}), we neglect the sub-leading term by taking $|r_3|=0$, which allows us to account for the correction originating from the weak fluctuations of charge in the left QD; see a third term in the Eq.~(\ref{result2}). At the resonance (quantum critical point), when $\mathcal{N}_g=1/2$ and reflection coefficients $|r_1|$, $|r_2|$ tuned in such a way to nullify constant $\mathcal{C}_+$ ($|r_+|=0$), this term is related to the leading irrelevant operator near the fixed point of the (two-channel) CK problem~\cite{furusakimatveev}. In this case, Eq.~(\ref{result2}) can be treated as a universal charge Kondo scaling 
\begin{eqnarray}\label{kondo}
G(T)=KG_0\left[1-\left(\frac{T}{T_K}\right)^{2K}\right]
\end{eqnarray}
with the Kondo temperature $T_K \sim E_C/|r_-|^{1/K}$ obtained in the case of weak backscattering $|r_1|,|r_2|$$\ll$$1$. At $T\gg T_K$, we would expect restoring of a logarithmic dependence of the conductance, $G(T)\propto\log^{-2}(T/T_K)$. However, it cannot be obtained for the considered in this paper model since $T\gg T_K$ at $|r_1|,|r_2|\ll 1$ requires $T>E_C$ condition. In this case, the CK model is not applicable due to more than just two charge states of the QD involved into consideration. In experiments, this problem is resolved by studying the regime of tunnel barriers, $1-|r_\alpha|^2\ll1$~~\cite{pierre1, pierre3}. 

The dependencies of Eq.~(\ref{kondo}) and the Kondo temperature on the LL interaction parameter $K=N_3/(2N_3+1)$ indicate the influence of the effective electron-electron interaction on the charge Kondo correlations. 
Interestingly, the Eq.~(\ref{kondo}) coincides with the temperature behavior of the conductance in the LL-based single-site two-channel charge Kondo (2CCK) problem first studied in Refs.~\cite{parafilo2,parafilo4}, see also \cite{parafilonew,karkinew}. The similarity becomes obvious (see, e.g., Eq.~(1) from~\cite{parafilo2} or Eq.~(12) from~\cite{parafilo4}) after presenting $K$ as $K=K_{red}/2$, where $K_{red}=2K_1K_2/(K_1+K_2)$ is a reduced Luttinger parameter of two LLs with different interaction constants, $K_1=1$ and $K_2=N_3/(N_3+1)$. Indeed, the right part of the device shown in Fig.~\ref{Fig2}(a) -- blue QD with all entered edge channels, represents by itself the LL simulator experimentally studied in Refs.~\cite{pierre,DCB}. Thus, we conclude that the 2SCK device with $N_1=1$ and $|r_3|=0$ [shown in Fig.~\ref{Fig2}(a)] at the special resonant point (when $r_+=0$) 
can be treated as the single-site 2CCK circuit, where one Kondo channel is attributed to non-interacting electron gas ($K_1=1$) and the second Kondo channel is attributed to the LL ($K_2=N_3/[N_3+1]$), see Fig.~\ref{Fig2}(b). Equation~(\ref{kondo}) in terms of new notation reads as $G(T)/G_0=(K_{red}/2)[1-(T/T_K)^{K_{red}}]$. 
Similarly, the first two terms in r.h.s. of Eq.~(\ref{result2}) written in terms of the reduced Luttinger parameter, $G(T)-K_{red}G_0/2\propto - (T/E_C)^{K_{red}-2}$, coincide with the conductance scaling of a resonant tunneling in the LL~~\cite{furusakinagaosa1}, see also~\cite{parafilo2,parafilo4}.

Similarly, the physics of the 2CCK problem, where each Kondo channel is implemented by the Luttinger liquid with its own interaction parameters $ K_1\neq1$ and $ K_2\neq1$, can be simulated in a hybrid metal-semiconductor device consisting of three islands. In this geometry, the middle QD is strongly coupled to the left (right) QD via single-mode QPC characterized by weak backscattering amplitude $r_L$($r_R$). This central island in the weak Coulomb blockade regime constitutes the Kondo impurity pseudo-spin. Meanwhile, the left (right) QD electrically connected to 2DEG via $N_1$($N_2$) fully open single-mode QPCs provides the LL simulator as in  Refs.~\cite{pierre,DCB}. In this case, we expect appearance of the conductance scaling Eq.~(\ref{kondo}) with $K=N_1N_2/(2N_1N_2+N_1+N_2)$ or in terms of the reduced Luttinger interaction parameter: $G(T)/G_0=(K_{red}/2)[1-(T/T_K)^{K_{red}}]$, where $K_{red}=2K_1K_2/(K_1+K_2)$ and $K_{\alpha}=N_{\alpha}/(N_{\alpha}+1)$ ($\alpha=1,2$). Corresponding calculation and the expression analog to Eq.~(\ref{result2}) will be done elsewhere. 

Moreover, a proposal to use a hybrid metal-semiconductor two- or three-island nanodevice for experimental study of the interaction effects on the 2CCK physics looks more promising than the original proposals in Refs.~\cite{parafilo2,parafilo4}. The main reason is that the 2SCK circuit is already fabricated \cite{DGG}. At the same time, a single-site analog of the setup was successfully used to study a universal conductor-insulator crossover in circuit quantum simulation of the LL with impurity~\cite{pierre,DCB}. Meanwhile, implementation of the model proposed in~\cite{parafilo2} encounters the obstacle associated with the fabrication of wide enough QPCs to observe strong enough electron-electron interaction between conducting channels. Another proposal from~\cite{parafilo4} is naturally constrained due to the problem of achieving a fractional quantum Hall regime with $\nu=1/3$ in an already fabricated device~\cite{pierre1, pierre3}.

\textit{Note.} While preparing the current manuscript, the author discovered a work~\cite{karkinew} addressing related questions in a similar hybrid metal-semiconductor double-quantum dot device but with a different configuration. Study in Ref.~\cite{karkinew}  concentrates on the conductance behavior near the quantum critical point in the case when $N$ ballistic channels connect two islands. As the main result, equation equivalent to Eq.~(\ref{result2}) of the current manuscript has been obtained. For the particular case of $N=2$, the appearance of the sub-leading temperature dependence with $g=3/5$ has been also reported in the case of finite backscattering in two QPCs placed between two islands.

We also urge the readers to investigate works \cite{thanh2018,kiselev,thanh2024,VN2}, which are closely related to this article and devoted to the study of the transport properties through the 2SCK device in the case of weak inter-dot coupling.

\section{Conclusions}

In this paper, we have theoretically investigated the low-temperature charge transport properties in a multi-channel two-site charge Kondo circuit, a device comprised of two large metallic islands (QDs) embedded into a high-mobility two-dimensional electron gas~\cite{DGG}. It is revised that the leading temperature behavior of the linear conductance obeys the conductance scaling of a single-impurity problem in the Luttinger liquid, $KG_0-G(T) \propto T^{2K-2}$, whose effective interaction parameter $K$ is fully determined by the number of conducting channels 
in the system, $K=N_1N_3/(N_1+N_3+N_1N_3)$ [here, $N_1$($N_3$) is the number of channels coupling left (right) QD to 2DEG].  We predict the appearance of extra sub-leading temperature dependencies $\propto T^{2g-2}$ ($\propto T^{2g'-2}$), where $g(g')=[N_1(N_3)-1+N_1N_3]/[N_1+N_3+N_1N_3]$ is attributed with a finite backscattering in $N_1$($N_3$) QPCs. In the particular case, when the left QD is electrically coupled to the source via single-mode QPC ($N_1=1$) and the right QD is electrically connected to 2DEG via $N_3$ fully open conducting channels, the 2SCK device can be treated as a single-site CK device with two channels associated with the non-interacting 1DEG and with the LL, respectively. The corresponding finite temperature correction in the vicinity of intermediate coupling 2CCK fixed point reads $KG_0-G(T)\propto (T/T_K)^{2K}$. Thus, one concludes that the 2SCK circuit could be used as a simulator to explore the interplay between interaction effects and multi-channel Kondo physics.

\textit{Acknowledgements.} The author would like to especially thank V. M. Kovalev and I. G. Savenko for fruitful discussions. The author also acknowledges support from the Institute for Basic Science in Korea (IBS-R024-D1).

\vspace*{5mm}

\begin{appendix}
\section{Diagonalization of the Hamiltonian}\label{appA}

In this Appendix, we discuss a unitary transformation, which diagonalizes the Hamiltonian $H_0+H_C$ and transfers bosonic fields 
$(\phi_{1,c},\phi_2,\phi_{3,c})^T$ to $(\phi_A,\phi_B,\phi_C)^T$. The corresponding transformation reads $H'_0+H'_C=\hat{P}^{-1}(H_0+H_C)\hat{P}$, where the matrix $\hat{P}$ reads 
\begin{eqnarray}\label{matrix}
\hat{P}=\left[\begin{array}{ccc}
p_{11} & p_{12}  & p_{13} \\ p_{21} & p_{22} & p_{23}  \\ p_{31} & p_{32} & p_{33}  
\end{array}\right],
\end{eqnarray}
with
\begin{eqnarray}
&&p_{11}=\sqrt{ \frac{K}{N_1}}\,,\,
p_{12}=\frac{\sqrt{N_1}(A-\Delta)}{D_1}\,,\,
 p_{13}=\frac{\sqrt{N_1}(A+\Delta)}{D_2},\nonumber\\ &&p_{21}=\sqrt{K}\,,\, p_{22}=-\frac{1+A-\Delta}{D_1}\,,\,  
p_{23}=-\frac{1+A+\Delta}{D_2},\nonumber\\
&& p_{31}=\sqrt{ \frac{K}{N_3}}\,,\, p_{32}=\frac{\sqrt{N_3}}{D_1}\,,\, p_{33}=\frac{\sqrt{N_3}}{D_2},
\end{eqnarray}
where we denote $A=(N_1-N_3)/2$, $\Delta=\sqrt{1+A^2}$ and $D_{1(2)}=\sqrt{N_3+(1+A\mp\Delta)^2+N_1(A\mp\Delta)^2}$.
%
%


\section{Effective Hamiltonians}\label{appB}

In this Appendix, we present explicitly the effective boundary sine-Gordon models, which appear after integrating out fluctuations of fields $\phi_B$, $\phi_C$ for particular cases of $N_1$,$N_3=\{1,2,3\}$.

\textbf{i) case}: $N_1=1$, $N_3=2$. Corresponding Luttinger interaction parameter and effective mass parameters of fields $\phi_B$ and $\phi_C$ read 
$K=2/5$, $\mathcal{M}_B=(5-\sqrt{5})/2$, $\mathcal{M}_C=(5+\sqrt{5})/2$, respectively. An effective Hamiltonian has the following form
\begin{widetext}
\begin{eqnarray}
&&H_{\rm eff}^{(i)}=\frac{v_F}{2\pi}\sum_{i=s,A}\int dx \left\{[\pi \Pi_i(x)]^2+[\partial_x\phi_i(x)]^2\right\}+\frac{D}{\pi}\left(\frac{\gamma E_C}{\pi D}\right)^{2/5}2|r_3|c_3\cos[\sqrt{2}\phi_s(0)]\cos\left[\sqrt{\frac{2}{5}}\phi_A(0)+\pi\mathcal{N}_g\right]\nonumber\\
&&~~~~~~~~~~~~~~+ \frac{D}{\pi}\left(\frac{\gamma E_C}{\pi D}\right)^{3/5}\left\{|r_1|c_1\cos\left[2\sqrt{\frac{2}{5}}\phi_A(0)-2\pi\mathcal{N}_g\right]\right.\left.+|r_2|c_2\cos\left[2\sqrt{\frac{2}{5}}\phi_A(0)\right]\right\},
\end{eqnarray}
\end{widetext}
where $c_1\approx 1.31$, $c_2\approx 2.$, $c_3\approx 1.38$.

{\bf ii) case}: $N_1=N_3=2$. Corresponding Luttinger interaction parameter and effective mass parameters of fields $\phi_B$ and $\phi_C$ read $K=1/2$, $\mathcal{M}_B=2$, $\mathcal{M}_C=4$, respectively. An effective Hamiltonian has the following form
\begin{widetext}
\begin{eqnarray}\label{model2}
&&H_{\rm eff}^{(ii)}=\frac{v_F}{2\pi}\sum_{i=s,A,\tilde{s}}\int dx \left\{[\pi \Pi_i(x)]^2+[\partial_x\phi_i(x)]^2\right\}+\frac{D}{\pi}\left(\frac{\gamma E_C}{\pi D}\right)^{1/2}|r_2|c_2\cos[\sqrt{2}\phi_A(0)]\\
&&~~~~~~~+ \frac{2D}{\pi}\left(\frac{\gamma E_C}{\pi D}\right)^{3/8}c_1\left\{|r_1|\cos[\sqrt{2}\phi_{s}(0)]\cos\left[\frac{1}{\sqrt{2}}\phi_A(0)-\pi \mathcal{N}_g\right]+|r_3|\cos[\sqrt{2}\phi_{\tilde{s}}(0)]\cos\left[\frac{1}{\sqrt{2}}\phi_A(0)+\pi \mathcal{N}_g\right]\right\},\nonumber
\end{eqnarray}
\end{widetext}
where $c_1=\sqrt{2}$, $c_2=2$. 

Interestingly, the above model is equivalent to the single-impurity problem in the Luttinger liquid with interaction parameter $K=1/2$ in case $|r_1|=|r_3|=0$. As it is well-known, the following model allows for an exact solution due to possibility of refermionization~\cite{gogolin,emerykivelson,giamarchi} of the problem and its further mapping on exactly solvable resonant model. Thus, the linear conductance can be exactly obtained from Eq.~(\ref{model2}) in case $|r_1|=|r_3|=0$, see~\cite{footnote}.

{\bf iii) case}: $N_1=1$, $N_3=3$. Corresponding Luttinger interaction parameter and effective mass parameters of fields $\phi_B$ and $\phi_C$ read $K=3/7$, $\mathcal{M}_B=3-\sqrt{2}$, $\mathcal{M}_C=3+\sqrt{2}$, respectively. An effective Hamiltonian has the following form
\begin{widetext}
\begin{eqnarray}
&&H_{\rm eff}^{(iii)}=\frac{v_F}{2\pi}\sum_{i=s,f,A}\int dx \left\{[\pi \Pi_i(x)]^2+[\partial_x\phi_i(x)]^2\right\}\\
&&~~~~~~+ \frac{D|r_1|c_1}{\pi}\left(\frac{\gamma E_C}{\pi D}\right)^{4/7} \cos\left[\frac{2\sqrt{3}}{\sqrt{7}}\phi_A(0)-2\pi \mathcal{N}_g\right] + \frac{D|r_2|c_2}{\pi}\left(\frac{\gamma E_C}{\pi D}\right)^{4/7} \cos\left[2\sqrt{\frac{3}{7}}\phi_A(0)\right]\nonumber\\
&&~+\frac{D|r_3|c_3}{\pi}\left(\frac{\gamma E_C}{\pi D}\right)^{\frac{2}{7}}\left\{\cos\left[\frac{2\phi_A(0)}{\sqrt{21}}-\frac{2\sqrt{2}}{\sqrt{3}}\phi_f(0)-\frac{2\pi\mathcal{N}_g}{3}\right]\right.\left.+2\cos\left[\frac{2}{\sqrt{21}}\phi_A(0)+\frac{\sqrt{2}}{\sqrt{3}}\phi_f(0)-\frac{2\pi\mathcal{N}_g}{3}\right]\cos\sqrt{2}\phi_{s}(0)\right\},\nonumber
\end{eqnarray}
where $c_1\approx1.346$, $c_2\approx1.93$, $c_3\approx1.39$.

{\bf iv) case}: $N_1=2$, $N_3=3$. Corresponding Luttinger interaction parameter and effective mass parameters of fields $\phi_B$ and $\phi_C$ read $K=6/11$, $\mathcal{M}_B=(7-\sqrt{5})/2$, $\mathcal{M}_C=(7+\sqrt{5})/2$, respectively. An effective Hamiltonian has the following form
\begin{eqnarray}
&& H^{\rm (iv)}_{\rm eff}=\frac{v_F}{2\pi}\sum_{i=s,\tilde{s},\tilde{f},A}\int dx \left\{[\pi \Pi_i(x)]^2+[\partial_x\phi_i(x)]^2\right\}\\
&& +\frac{D|r_2|c_2}{\pi}\left(\frac{\gamma E_C}{\pi D}\right)^{\frac{5}{11}}\cos\left[2\sqrt{\frac{6}{11}}\phi_A(0)\right]+\frac{2D|r_1|c_1}{\pi}\left(\frac{\gamma E_C}{\pi D}\right)^{\frac{4}{11}}\cos\sqrt{2}\phi_s(0)\cos\left[\sqrt{\frac{6}{11}}\phi_A(0)-\pi\mathcal{N}_g\right]\nonumber\\
&&+\frac{D|r_3|c_3}{\pi}\left(\frac{\gamma E_C}{\pi D}\right)^{\frac{3}{11}}\left\{\cos\left[\sqrt{\frac{8}{33}}\phi_A(0)-\frac{2\sqrt{2}}{\sqrt{3}}\phi_{\tilde{f}}(0)+\frac{2\pi\mathcal{N}_g}{3}\right]+2\cos\left[\sqrt{\frac{8}{33}}\phi_A(0)+\sqrt{\frac{2}{3}}\phi_{\tilde {f}}(0)+\frac{2\pi\mathcal{N}_g}{3}\right]\cos\sqrt{2}\phi_{\tilde{s}}(0) \right\},\nonumber
\end{eqnarray}
where $c_1\approx1.426$, $c_2\approx1.946$, $c_1\approx1.405$.

{\bf v) case}: $N_1=3$, $N_3=3$. Corresponding Luttinger interaction parameter and effective mass parameters of fields $\phi_B$ and $\phi_C$ read $K=6/11$, $\mathcal{M}_B=3$, $\mathcal{M}_C=5$, respectively. An effective Hamiltonian has the following form
\begin{eqnarray}
&& H^{\rm (v)}_{\rm eff}=\frac{v_F}{2\pi}\sum_{i=s,\tilde{s},A,f,\tilde{f}}\int dx \left\{[\pi \Pi_i(x)]^2+[\partial_x\phi_i(x)]^2\right\}+\frac{D|r_2|c_2}{\pi}\left(\frac{\gamma E_C}{\pi D}\right)^{2/5}\cos\left[2\sqrt{\frac{3}{5}}\phi_A(0)\right]\\
&&+\frac{D|r_1|c_1}{\pi}\left(\frac{\gamma E_C}{\pi D}\right)^{\frac{4}{15}}\left\{\cos\left[\frac{2\phi_A(0)}{\sqrt{15}}-\frac{2\sqrt{2}}{\sqrt{3}}\phi_f(0)-\frac{2\pi\mathcal{N}_g}{3}\right]+2\cos\left[\frac{2\phi_A(0)}{\sqrt{15}}+\frac{\sqrt{2}}{\sqrt{3}}\phi_f(0)-\frac{2\pi\mathcal{N}_g}{3}\right]\cos\sqrt{2}\phi_s(0)\right\}\nonumber\\
&& +\frac{D|r_3|c_3}{\pi}\left(\frac{\gamma E_C}{\pi D}\right)^{\frac{4}{15}}\left\{\cos\left[\frac{2\phi_A(0)}{\sqrt{15}}-\frac{2\sqrt{2}}{\sqrt{3}}\phi_{\tilde{f}}(0)+\frac{2\pi\mathcal{N}_g}{3}\right]+2\cos\left[\frac{2\phi_A(0)}{\sqrt{15}}+\frac{\sqrt{2}}{\sqrt{3}}\phi_{\tilde{f}}(0)-\frac{2\pi\mathcal{N}_g}{3}\right]\cos\sqrt{2}\phi_{\tilde{s}}(0)\right\},\nonumber
\end{eqnarray}
where $c_2\approx1.9$, $c_1=c_3\approx1.41$.

\end{widetext}

\end{appendix}

\newpage


\begin{thebibliography}{10}
\bibitem{footnote}
In case $N_1=N_3=2$ ($K=1/2$) and $|r_1|=|r_3|=0$, the problem allows for an exact solution~\cite{furusakimatveev,emerykivelson,gogolin}. The exact expression for the conductance reads $G=KG_0[1-\int d\epsilon (-\partial f(\epsilon)/ \partial \epsilon) \Gamma_0^2/(\epsilon^2+\Gamma_0^2)]$, where $f(\epsilon)$ is the Fermi-Dirac distribution function and $\Gamma_0=(4\gamma E_C/\pi^2)|r_2|^2$.
\end{thebibliography}
\end{document}